\documentclass[11pt,amssymb,amsmath,amssymb,floatfix]{article}
\usepackage[T1]{fontenc}
\usepackage[utf8]{inputenc}
\usepackage{geometry}
\usepackage{textcomp}
\usepackage{mathrsfs}
\usepackage{amsmath}
\usepackage{graphicx}
\usepackage{subfig}
\usepackage{graphicx}
\usepackage[colorlinks=true]{hyperref}
\usepackage[numbers,sort&compress]{natbib}
\usepackage{slashed}
\usepackage{color}

\makeatletter

\newcommand{\lyxmathsym}[1]{\ifmmode\begingroup\def\b@ld{bold}
	\text{\ifx\math@version\b@ld\bfseries\fi#1}\endgroup\else#1\fi}

\makeatother

\begin{document}
	\title{\Large{\bf{Enhanced Quark-Loop Contribution to Pure Annihilation Nonleptonic B-meson decays in PQCD approach }}}
	
	\maketitle
	\begin{center}       
		
		{ Hong-Yuan Sheng$^{a}$, 
			Feng-Chen Chu$^{a}$, 
			Yue-Long Shen$^{a,b,c}$\footnote{Corresponding author: shenylmeteor@ouc.edu.cn}, 
			Zhi-Tian Zou$^d$}\\     
		\vskip0.5cm         
		{ \it $^a$  College of Physics and Opto-electric Engineering, Ocean   University of China, Qingdao 266100,  China } \\   
		{\it  $^b$ Engineering Research Center of Advanced Marine Physical Instruments and Equipment, Ministry of Education, Qingdao 266100,  China }\\
		{\it  $^c$Qingdao Key Laboratory of Optics and Optoelectronics, Qingdao 266100,  China  } \\
		{ \it $^d$ Department of Physics, Yantai University, Yantai 264005, China} \\         
	\end{center}
	\vskip1.3cm
	\begin{abstract}
		  In this study, we conduct an enhanced investigation of pure annihilation-type charmless hadronic B-decays within the framework of the PQCD approach. Specifically, we account for the quark-loop enhanced contribution to various decay processes at the next-to-leading order(NLO) in the strong coupling constant $\alpha_s$.     The NLO amplitudes possess large imaginary parts that have signs opposite to those of the leading-order amplitudes. This cancellation effect implies that the NLO contribution does not significantly influence the branching ratios of the processes under consideration. However, the NLO effect explored in this work constitutes an important source of the strong phases in the weak - annihilation non-leptonic $\bar B_q$-meson decay amplitudes. As a result, it can have a substantial impact on CP - violating observables such as the direct CP asymmetry ${\cal A}_{\rm CP}^{\rm dir}$ and mixing induced CP asymmetry ${\cal A}_{\rm CP}^{\rm mix}$. The numerical result evidently revealed that  NLO QCD correction significantly enhances both  ${\cal A}_{\rm CP}^{\rm dir}$ and ${\cal A}_{\rm CP}^{\rm mix}$ of the pure annihilation type B decay processes.
		\newpage{}
	\end{abstract}
	
	\section{Introduction}
	Exclusive two-body charmless bottom-meson decays hold fundamental significance. They are crucial for determining the Cabibbo - Kobayashi - Maskawa (CKM) matrix elements and exploring CP violation in electroweak interactions. Moreover, these decays offer an ideal context for enhancing our understanding of the diverse aspects of strong interaction dynamics that govern flavor-changing heavy-quark decay processes.  Approximately forty years ago, the factorization approach based on the color transparency assumption was employed to study hadronic B - decay processes. However, it failed to accurately predict experimental observables such as CP violations. In the past two decades,  the development of QCD-inspired effective theories has remarkably enhanced the accuracy of theoretical predictions for the partial widths of these decays.
	The QCD factorization (QCDF) \cite{Beneke:1999br,Beneke:2000ry} approach, founded on collinear factorization, has been extensively utilized to study various two-body nonleptonic B - decay channels,  and most of its predictions for decay rates are consistent with experimental measurements. The QCD factorization framework can be more gracefully reformulated using the language of the soft - collinear effective theory (SCET) \cite{Chay:2003ju,Beneke:2002ph,Bauer:2004tj,Bauer:2005kd}.
	The Perturbative QCD (PQCD) \cite{Keum:2000wi,Lu:2000em,Li:2004ep,Ali:2007ff,Li:2010nn,Li:2012nk,Li:2012md,Li:2013xna,Li:2014xda} approach is established on the basis of transverse-momentum-dependent factorization. Significantly, within this framework, the CP-violation observables of most nonleptonic B-decay processes have been successfully predicted.
	
	The QCDF approach performs effectively at leading power in the heavy quark expansion. However,  power suppressed contributions, like those from the annihilation topology, cannot be accurately computed. This is because of the troublesome endpoint divergences present in the convolution integrals of  perturbation function and the light-cone distribution amplitudes of  initial and final states.  Although efforts have been made to address the endpoint singularity  within the framework of collinear factorization, a universally accepted solution remains elusive \cite{Liu:2019oav,Liu:2020tzd,Liu:2020wbn,Beneke:2022obx,Bell:2022ott}.   In the QCDF approach, the annihilation contribution is estimated through process-independent parametrizations for logarithmically and linearly divergent integrals \cite{Beneke:2001ev,Beneke:2003zv,Beneke:2006hg,Bartsch:2008ps} (see also \cite{Cheng:2008gxa,Cheng:2009cn,Cheng:2009mu,Zhu:2011mm,Wang:2013fya,Bobeth:2014rra,Chang:2014yma,Chang:2016qyc,Chang:2017brr} and the references therein), unfortunately, this methods leads to substantial theoretical uncertainties.   In contrast, within the framework of the PQCD approach, the endpoint singularity is regularized by taking into account the transverse momentum of the partons within the initial and final state mesons. As a result, the annihilation diagrams are considered factorizable in the PQCD approach. It has been discovered that this contribution can supply the strong phases necessary for direct CP asymmetries in nonleptonic B meson decays.

	Pure annihilation-type nonleptonic B-meson decays can only take place via annihilation diagrams within the Standard Model (SM). This is because none of the quarks (or anti-quarks) in the final states match those of the initial B meson. Studying such processes offers a unique perspective on understanding the QCD dynamics in heavy  hadron decays.
	Within the PQCD framework, various pure annihilation-type B decays have been explored at the $\alpha_s$  order \cite{Li:2004ep,Lu:2001yz,Lu:2002iv,Li:2003ku,Li:2003az,Song:2004cd,Li:2005vu,Lu:2005be,Zhu:2005rt,Li:2006xe,Rui:2011qc,Yu:2005rm,Yang:2010ba,Xiao:2011tx}. The predicted branching ratios are in agreement with experimental measurements. With the continuous accumulation of experimental data, more observables, such as CP asymmetries  in the pure annhilation-type B decays, are expected to be measured in the near future.
	For certain decay modes like  $B_s\to \pi\pi$,  the predicted CP violation was very small in the previous studies. This is because the direct CP violation is proportional to the ratio of the magnitudes of the contributions from the tree operators and the penguin operators, as well as the sine of the strong phase difference between them. Moreover, the contribution from the tree operators is suppressed by the CKM mechanism compared with that of the penguin operators, and the strong phase difference between them is also very small.   A recent study \cite{Lu:2022kos} has revealed that the charm-quark loop contribution can effectively modify the ratio of the magnitudes  and strong phase difference of the contributions from the tree operators and the penguin operators, despite an additional $\alpha_s$ suppression. It has been found that the quark loop contribution has a significant impact on CP asymmetries, while only slightly affecting the branching ratios.
	In \cite{Lu:2022kos}, to ensure the convergence of the convolution integral, only the contributions from twist-2 LCDAs  of light mesons were considered. As a result, the predicted branching ratios for pure annihilation charmless hadronic B decays are considerably smaller than experimental measurements.

	In this paper, within the framework of the PQCD approach, we will conduct a more in-depth study of the pure annihilation type charmless hadronic B-meson decays, which include processes such as $B^0\to K^+K^-,$ $ K^{\ast+}K^{\ast-},\phi\phi$  and $B_s\to \pi\pi, \rho\rho, \rho\omega,\omega\omega$ .     The main improvement of this paper lies in taking into account the quark loop contributions in various decay channels, and this contribution is expected to have a significant impact on the CP violation observables.  Since the PQCD approach can effectively eliminate the endpoint singularities, in our calculations, we will incorporate the contributions from the twist-2 and two-particle twist-3 light-cone distribution amplitudes of the final-state light mesons. It is expected that, compared with the leading-order contributions in previous studies, the introduction of the quark loop does not significantly change the branching ratios of the pure annihilation decay processes, while the observables of the direct CP violation and the mixing-induced CP violation  have both changed significantly.  In this paper, we do not consider the $B \to PV$($P$ denotes a pseudo-scalar meson and $V$ stands for a vector meson) decays, because the quark loop effect does not contribute to these processes. The structure of this paper is arranged as follows. In the next section, we will briefly review the theoretical framework of the PQCD  factorization approach and the calculation of the leading-order decay amplitude. Then, in Section 3, we will perform the perturbative calculations for the enhanced quark loop contributions, while the numerical results and phenomenological analysis will be presented in Section 4. In the last section, we will provide a brief summary and discussion.
	
	\section{The decay amplitudes at leading order}
	
	We start from the effective weak Hamiltonian
	of the non-leptonic $\Delta B =1$ transitions in the Standard Model
	\begin{eqnarray}
		{\cal H}_{\rm eff} ={ G_F \over \sqrt{2}} \,
		\sum_{p=u, c} V_{p b} \, V_{p q}^{\ast}
		\bigg [ C_1(\nu) \, Q_1^{p}(\nu)  + C_2(\nu) \, Q_2^{p}(\nu)
		+ \sum_{i=3}^{10} C_i(\nu) \, Q_i(\nu) + C_{8 g}(\nu) \,\, Q_{8 g}(\nu)   \bigg  ] \,,
	\end{eqnarray}
	where we adopt the effective operator basis  presented in
	\cite{Buchalla:1995vs} . The corresponding Wilson coefficients $C_i(\nu)$ with $\nu \sim {\cal O}(m_b)$  can be derived through renormalization group evolution.
	The decay amplitude of $B \to M_2M_3$ decays is, in fact, the matrix elements of the effective Hamiltonian, i.e.,
	\begin{eqnarray}
		\bar {\cal A} (\bar B_q \to M_2 M_3)
		=  \langle M_2 (p_2)  \, M_3 (p_3)  | {\cal H}_{\rm eff}  | \bar  B_q (p_B) \rangle.
	\end{eqnarray}
	
In the PQCD approach, the decay amplitudes are free from endpoint singularities. This is because the transverse momenta of the partons within the initial- and final- state hadrons are retained, and these amplitudes are considered to be completely factorizable. The typical factorization formula in the PQCD approach is given by:
	\begin{eqnarray}
		{\cal M}&\propto&\int^1_0\prod_{i=1}^3 dx_i\int\prod_{j=1}^3 
		{d^2{\vec{b}}_{j}}{\phi}_B(x_1,{\vec{b}}_{1},t){\phi}_{M_2}(x_2,{\vec{b}}_{2},t){\phi}_{M_3}(x_3,{\vec{b}}_{3},t)\nonumber\\
		&&\times
		T_H(x_1,x_2,x_3,{\vec{b}}_{1},{\vec{b}}_{2},{\vec{b}}_{3},t)
		\exp[-S_B(t)-S_2(t)-S_3(t)]~,
	\end{eqnarray}
	where $\vec{b}_{j}$ is the impact parameter  conjugated to the parton's  transverse momentum $\vec{k}_{\perp j}$.  The wave functions ${\phi}_{M}(x_i,{\vec{b}}_{i},t)$  are related to the  transverse-momentum-dependent(TMD) wave functions ${\phi}_{M}(x_i,{\vec{k}}_{\perp i},t)$ via Fourier transformations.   The TMD wave functions defined on the lightcone possess light-cone singularity, which needs to be eliminated in a more precise definition.    It is convenient to sum up the large logarithms $\ln^2 (k_\perp^2/Q^2)$ and $\ln (k_\perp^2/Q^2)$ ($Q^2$ denotes a hard scale) arising from the  radiative corrections to the wave functions in the impact parameter space. Resummation of the large   logarithms leads to the Sudakov form factors $S_B(t)$, $S_2(t)$ and $S_3(t)$, which are given in Appendix \ref{pqcd}. The perturbative function $T_H(x_1,x_2,x_3,{\vec{b}}_{1},{\vec{b}}_{2},{\vec{b}}_{3},t)$ is calculated in the momentum space,  and  it is also required to be transformed into  the impact space.  
	
	\begin{figure}
		\begin{center}
			\hspace{0cm}\scalebox{1}[1]{\includegraphics[width=0.5\textwidth]{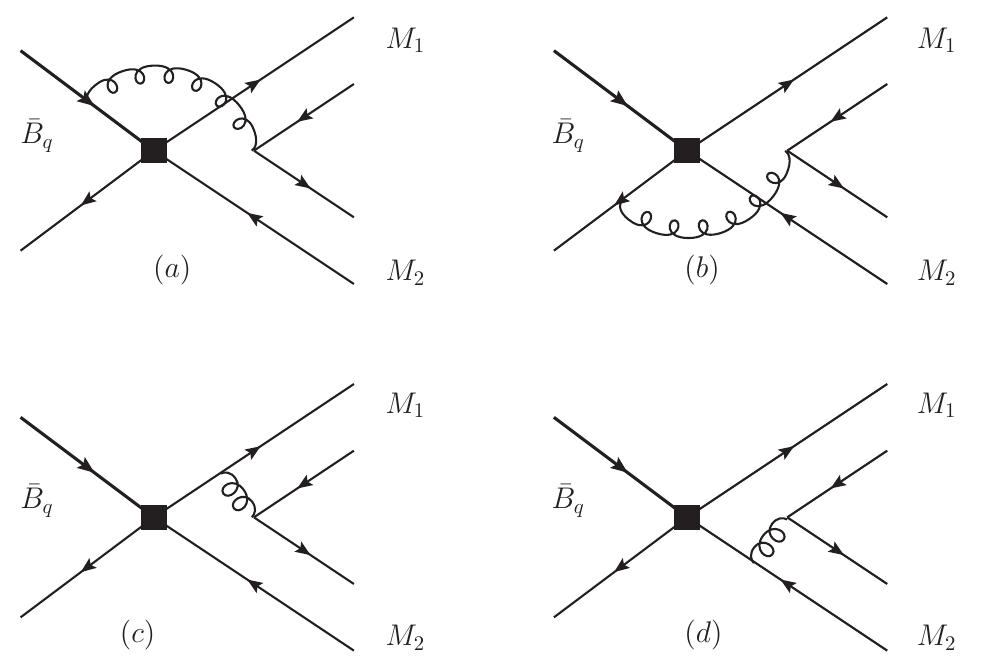}}
			\caption{ The leading order Feynman diagrams for annihilation
				contribution, with possible four-quark operator
				insertions. }\label{fig:LO}
		\end{center}
	\end{figure}
	The leading order Feynman diagrams for the pure annihilation-type nonleptonic  B decays  are depicted in Fig.\ref{fig:LO}. As a tree-level calculation,  the decay amplitudes can be computed using the trace formalism.  For final state light mesons, we disregard the parton's transverse momentum in the wave functions since it has negligible impact on the phenomenological results.    The  factorization formulae of the  $B^0\to K^+K^-,$ $ K^{\ast+}K^{\ast-},\phi\phi$  and $B_s\to \pi\pi, \rho\rho, \rho\omega,\omega\omega$ decays have already given in \cite{Ali:2007ff, Chai:2022ptk}.  In the following we present the specific expressions.  For the $B \to PP$ decays, we have 
	\begin{eqnarray}
		\mathcal{A}_{\rm LO}(\bar{B}^0_s\rightarrow \pi^+\pi^-)&=&-\frac{G_F}{\sqrt{2}}V_{tb}V^*_{ts}\left\{\mathcal{M}^{LL}_{\rm LO}\left[2C_4+\frac{1}{2}C_{10}\right]+\mathcal{M}^{SP}_{\rm LO}\left[2C_6+\frac{1}{2}C_{8}\right]\right\}
		\nonumber \\ 	&+&\frac{G_F}{\sqrt{2}}V_{ub}V^*_{us}\mathcal{M}^{SP}_{\rm LO}[C_2];\nonumber  \\
		\sqrt{2}\mathcal{A}_{\rm LO}(\bar{B}^0_s\rightarrow \pi^0\pi^0)&=&	\mathcal{A}_{\rm LO}(\bar{B}^0_s\rightarrow \pi^+\pi^-);\nonumber  \\
		\mathcal{A}_{\rm LO}(\bar{B}^0_d\rightarrow K^+K^-)&=&-\frac{G_F}{\sqrt{2}}V_{tb}V^*_{td}\left\{\mathcal{M}^{LL}_{\rm LO}\left[2C_4+\frac{1}{2}C_{10}\right]+\mathcal{M}^{SP}_{\rm LO}\left[2C_6+\frac{1}{2}C_{8}\right]\right\}
		\nonumber \\ 	&+&\frac{G_F}{\sqrt{2}}V_{ub}V^*_{ud}\mathcal{M}^{LL}_{\rm LO}[C_2],
	\end{eqnarray}
	where $\mathcal{M}^{LL(R)}_{\rm LO}$ and $\mathcal{M}^{ SP}_{\rm LO}$ are derived through calculating   the Feynman diagrams in Fig.\ref{fig:LO} (c,d)  by inserting four-quark operators   $\bar q_1\gamma^\mu(1-\gamma_5)b \bar q_2\gamma_\mu(1-(+)\gamma_5)q_1 $  and $\bar q_1(1-\gamma_5)b \bar q_2(1+\gamma_5)q_1 $ (after Fierz transformation) respectively, and they read
	\begin{align}
		\mathcal{M}^{LL}_{\rm LO}[C_i]&=32\pi C_FM^4_{B_s}/\sqrt{6}\int_{0}^{1}  \,dx_1 dx_2 dx_3 \int_{0}^{\infty }  \,b_1 db_1 b_2 db_2 
		\phi _{B_s}(x_1,b_1) \nonumber\\ 
		&
		\big\{\big[-x_2\phi^A_2\phi^A_3 -4r_2r_3\phi ^P_2\phi^P_3+r_2r_3(1-x_2)(\phi^P_2 +\phi^T_2)(\phi^P_3-\phi^T_3)\nonumber\\
		&+r_2r_3x_3(\phi^P_2 -\phi^T_2)(\phi^P_3+\phi^T_3)\big]a_i(t_d)E'_a(t_d) h_{na}(x_1,x_2,x_3,b_1 ,b_2)\nonumber\\
		&+ [(1-x_3)\phi^A_2\phi^A_3+r_2r_3(1-x_3)(\phi^P_2 +\phi^T_2)(\phi^P_3-\phi^T_3)\nonumber\\
		&+r_2r_3x_2(\phi^P_2 -\phi^T_2)(\phi^P_3+\phi^T_3)]C_i(t'_d)E'_a(t'_d)h'_{na}(x_1,x_2,x_3,b_1 ,b_2)\big\},
	\end{align}
	\begin{align}
		\mathcal{M}^{SP}_{ann}[C_i]&=32\pi C_FM^4_{B_s}/\sqrt{6}\int_{0}^{1}  \,dx_1 dx_2 dx_3 \int_{0}^{\infty }  \,b_1 db_1 b_2 db_2 
		\phi _{B_s}(x_1,b_1) \nonumber\\ 	&
		\{ [(x_3-1)\phi^A_2\phi^A_3 -4r_2r_3\phi ^P_2\phi^P_3+r_2r_3x_3(\phi^P_2 +\phi^T_2)(\phi^P_3-\phi^T_3)\nonumber\\
		&+r_2r_3(1-x_2)(\phi^P_2 -\phi^T_2)(\phi^P_3+\phi^T_3)]a_i(t_d)E'_a(t_d)h_{na}(x_1,x_2,x_3,b_1 ,b_2)\nonumber\\
		&+ [x_2\phi^A_2\phi^A_3+r_2r_3x_2(\phi^P_2 +\phi^T_2)(\phi^P_3-\phi^T_3)\nonumber\\
		&+r_2r_3(1-x_3)(\phi^P_2 -\phi^T_2)(\phi^P_3+\phi^T_3)]C_i(t'_d)E'_a(t'_d)h'_{na}(x_1,x_2,x_3,b_1 ,b_2)\}.
	\end{align}
	In the above equations  we have employed the shorthand notation $\phi ^{A,P,T}_i\equiv \phi ^{A,P,T}(x_i)$.  Here $\phi ^{A}$ represents   twist-2 LCDA of a pseudo-scalar meson, and   $\phi ^{P, T}$  stand for two-particle twist-3 LCDAs of a pseudo-scalar meson.  The ratios $r_i=m_{0i}/m_{B_q}$, where $m_{0i}$ is the chiral scale. 
	The contributions from Fig. \ref{fig:LO} (a) and (b) cancel each other out. This cancellation occurs due to the conservation of the vector current and axial-vector current, which are constructed from massless quark fields. 
	
	The $B \to VV$  decays  are more complicated than $B \to PP$ modes since they contain three kinds of polarizations states of final state vector meson,
	namely longitudinal (L), normal (N) and transverse (T) polarization.  The amplitude which are corresponding to the longitudinal, normal and transverse polarization amplitudes respectively $\mathcal{ A}^{(\lambda)}$
	can be decomposed as follows:
	\begin{eqnarray}
		\mathcal{A}(B_q\to VV) 	&=&\mathcal{A}_{L}+\mathcal{A}_{N}
		\epsilon^{*}_{2T}\cdot\epsilon^{*}_{3T} +i
		\frac{\mathcal{A}_{T}}{m_{B_q}^2}\epsilon^{\alpha \beta\rho\sigma}
		\epsilon^{*}_{2\alpha}\epsilon^{*}_{3\beta}
		p_{2\rho }p_{3\sigma }\;.
	\end{eqnarray}
	Similar to the $B \to PP$ decays,  the  amplitudes $\mathcal{A}_{L,N,T}$ can also be obtained by calculated the Feynman diagrams depicted in  Fig.\ref{fig:LO}. The specific expressions for the longitudinal, normal and transverse polarization amplitudes at leading order are collected in the appendix \ref{vvamp}. 
	\section{The quark loop enhanced contribution}
	\begin{figure}
		\begin{center}
			\hspace{0cm}\scalebox{1}[1]{\includegraphics[width=0.5\textwidth]{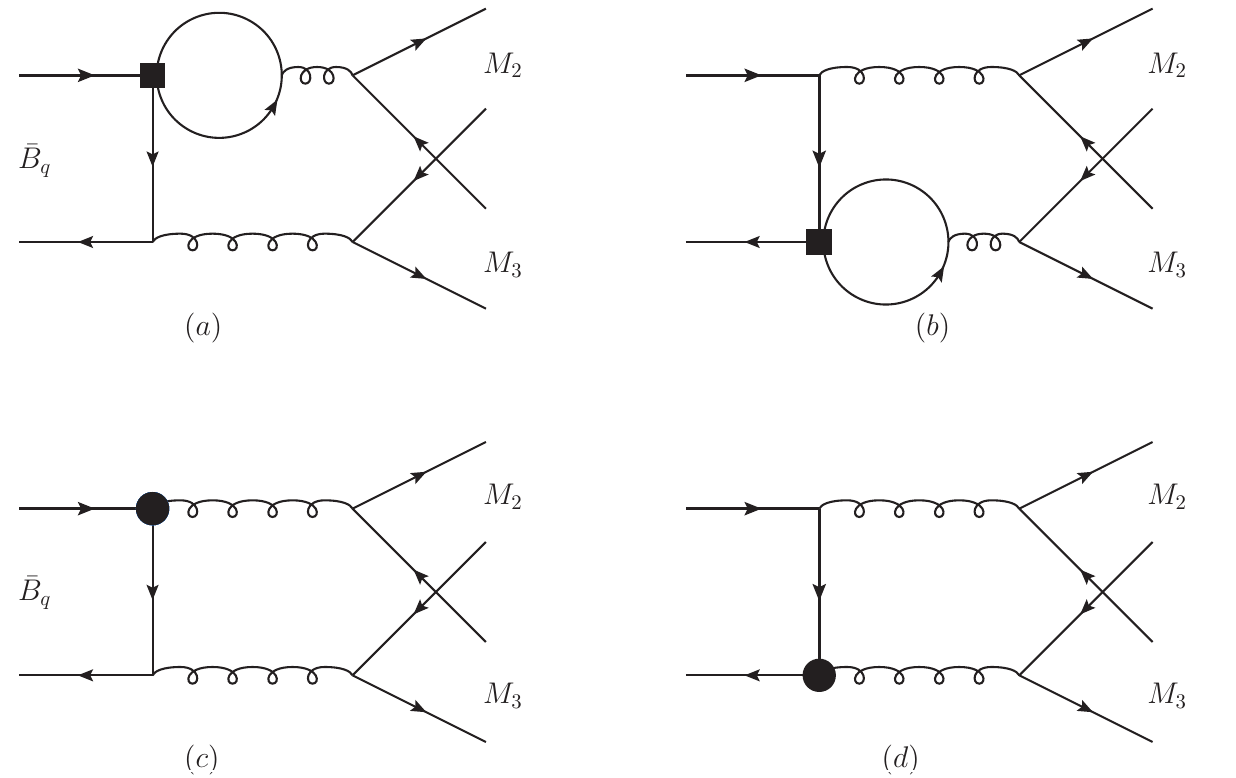}}
			\caption{ The Feynman diagrams for next-to-leading order corrections to annihilation contribution. The black square denotes four quark operators, and the black dot denotes the chromo-magnetic operator $O_{8g}$.  There also exist four diagrams where the gluon propagators  attached to $M_2$ and $M_3$ exchange with each other,   and they are not explicitly shown.  }\label{fig:NLO}
		\end{center}
	\end{figure}
	\begin{figure}
		\begin{center}
			\hspace{0cm}\scalebox{1}[1]{\includegraphics[width=0.35\textwidth]{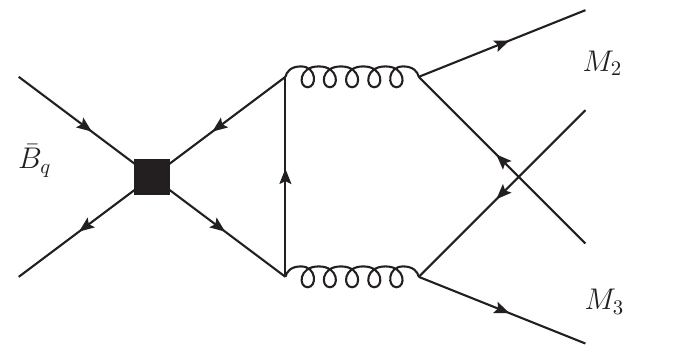}}
			\caption{ The triangle diagrams for next-to-leading order corrections to annihilation contribution. The black square denotes four quark operators.  The gluon propagator exchanged diagram is also not explicitly shown.}\label{triangle}
		\end{center}
	\end{figure}
	In this section, our objective is to compute the next-to-leading order (NLO) corrections to the decay amplitudes of pure annihilation non-leptonic B decays. We will focus on the contributions from quark loop diagrams and the chromo-magnetic penguin operator. The corresponding Feynman diagrams are shown in Fig. \ref{fig:NLO} and Fig. \ref{triangle}.
	It is true that there are other NLO corrections, such as QCD corrections to the diagrams in Fig. \ref{fig:LO}. However, these are less significant compared to the contributions considered in this work. This is because the quark loop diagrams may be enhanced by the CKM matrix elements or Wilson coefficients.  We take $\bar B_s \to \pi^+\pi^-$  decay as an example.   At leading order, this process is penguin dominated since the tree diagram contribution  is suppressed by  $|V_{ub}V_{us}^\ast/V_{tb}V_{ts}^\ast|\sim 0.02$, meanwhile,  the charm loop diagrams have both large Wilson coefficient $C_2$ and large CKM matrix element   $V_{cb}V_{cs}^\ast$ albeit with an additional factor $\alpha_s/4\pi$,  and  it is expected to be comparable to  the  leading order penguin contribution.  Upon careful examination\cite{Lu:2022kos}, it has been determined that the contribution from the triangle diagrams shown in Fig. \ref{triangle} is significantly more suppressed compared to that of the diagrams in Fig. \ref{fig:NLO}. As a result, these triangle-diagram contributions have been omitted from the current study for the sake of computational efficiency and theoretical focus. 
	
	The final states of the pure annihilation type charmless nonleptonic  B decays  are particle-antiparticle pair,   and it is obvious from Fig. \ref{fig:NLO}  that  NLO amplitude  of $B \to M_1 M_2$ decays can be expressed as 
	\begin{eqnarray}
		\mathcal{A}_{\rm NLO} 	&=&\langle M_2(p_2)M_3(p_3)|g^\ast(p_g,\alpha) g^\ast(\tilde p_g,\beta)\rangle\langle g^\ast(p_g,\alpha) g^\ast(\tilde p_g,\beta)|\mathcal{H}_{\rm eff}| \bar B_q\rangle.
	\end{eqnarray}
	At leading power, the  internal gluons must be transversely polarized since it attaches a collinear quark field and an anti-collinear quark field,   then the $\bar B_q \to g^\ast(p_g,\alpha) g^\ast(\tilde p_g,\beta)$  transition matrix element can be parameterized as
	\begin{eqnarray}
		&& \langle g^{\ast}(p_g, \alpha) \,  g^{\ast}(\tilde{p}_g, \beta) |{\cal H}_{\rm eff}  | \bar B_{q} \rangle
		= i \, \epsilon_{\alpha \beta p_g \tilde{p}_g} \, F_V(p_g^2, \, \tilde{p}_g^2)
		+ g_{\alpha \beta}^{\perp} \, \, F_A(p_g^2, \, \tilde{p}_g^2).
		\label{B to gg matrix element}
	\end{eqnarray}
	According to the analysis presented in \cite{Lu:2022kos} ,
	the form-factors $ F_{V,A}$  satisfy the  relations $F_{V}(p_g^2, \, \tilde{p}_g^2)= - F_{V}(\tilde{p}_g^2, \, p_g^2)$
	and $F_A(p_g^2, \, \tilde{p}_g^2)= F_A(\tilde{p}_g^2, \, p_g^2)$.
	The anti-symmetric property of $F_{V}(p_g^2, \, \tilde{p}_g^2)$ implies that the NLO contribution considered in this paper will not contribute to $B \to PV$ decays at leading power.  However, if twist-3 LCDAs are taken into account, the internal gluons may not be transverse,  this is because the Lorentz structure of a power suppressed collinear quark field  diffesr from that from leading power one, and the previous conclusion no longer be valid. Meanwhile,  the twist-3 contribution to  $\langle P(p_2)V(p_3)|g^\ast(p_g,\alpha) g^\ast(\tilde p_g,\beta)\rangle$ indeed vanishes,  because  the number of  gamma matrices of  the LCDAs pseudo-scalar and vectors mesons is odd.
	
	In the calculation of quark loop diagrams in Fig. \ref{fig:NLO}, we neglect the transverse momenta of the partons in  initial and final state mesons since  no additional endpoint divergence will be introduced, and a similar approach has been adopted in \cite{ Chai:2022ptk}.  In order to facilitate the comparison between the NLO results and the leading order results,  we firstly integrate out the quark fields in the loop and the gluon field attaching the quark loop and the $O_{8g}$ operator,  subsequently, we obtain two effective operators  
	\begin{align}
		C_{D,1}^{\rm eff}\bar D \gamma^\mu(1-\gamma_5)T^ab \bar q\gamma_\mu T^a q+
		C_{D,2}^{\rm eff}\bar D \not\! q \gamma^\mu(1+\gamma_5)T^ab \bar q\gamma_\mu T^a q, \label{effoper}
	\end{align}
	where $D=d, s$ denoting a down-type quark. The effective Wilson coefficients are expressed as
	\begin{align}
		C_{D,1}^{\rm eff}&=[V_{ub}V_{uD}^\ast G(0)+V_{cb}V_{cD}^\ast G(m_c)]C_2-V_{tb}V_{tD}^\ast [G(0)+G(m_b)]C_3   \nonumber \\ &-(C_4+C_6)V_{tb}V_{tD}^\ast[3G(0)+G(m_c)+G(m_b)-20/3],
		\nonumber \\
		C_{D,2}^{\rm eff}&=-{4m_b\over q^2}V_{tb}V_{tD}^\ast(C_5+C_{8g}).
	\end{align}
	The function $G(m)$ arising from the loop diagrams can be decomposed as 
	\begin{align}
		G(m)
		=G_q(m)+G_m(m),
	\end{align}
	with 
	\begin{align}
		&G_q(m)
		=-\frac{4}{3}\left(\frac{1}{\epsilon }+\ln\frac{\mu^2}{m^2}+\frac{1}{6}\right)+\frac{8}{3}\sqrt{\frac{4m^2}{q^2}-1}\arctan\frac{1}{\sqrt{\frac{4m^2}{q^2}-1}},
		\nonumber \\	
		&G_m(m)
		={16 m^2\over 3q^2}\sqrt{\frac{4m^2}{q^2}-1}\arctan\frac{1}{\sqrt{\frac{4m^2}{q^2}-1}}-{28m^2\over 3q^2}.
	\end{align}
	If $m\to 0$, we have 
	\begin{align}
		&G_0=G_q(0)
		=-\frac{4}{3}\left[\frac{1}{\epsilon }+\ln\left(-\frac{\mu^2}{q^2}\right)+\frac{1}{6}\right].
	\end{align}
	Accordingly,  the effective Wilson coefficient $C_{D,1}^{\rm eff}$  needs to be written as $C_{D,1}^{\rm eff}=C_{D,1q}^{\rm eff} +C_{D,1m}^{\rm eff} $,   here $C_{D,1q}^{\rm eff}$ and $C_{D,1m}^{\rm eff}$  have the following form 
	\begin{align}
		C_{D,1q}^{\rm eff}&=[V_{ub}V_{uD}^\ast G_0+V_{cb}V_{cD}^\ast G_q(m_c)]C_2-V_{tb}V_{tD}^\ast [G_0+G_q(m_b)]C_3   \nonumber \\ &-(C_4+C_6)V_{tb}V_{tD}^\ast[3G_0+G_q(m_c)+G_q(m_b)-20/3],
		\nonumber \\
		C_{D,1m}^{\rm eff}&=V_{cb}V_{cD}^\ast G_m(m_c)C_2-V_{tb}V_{tD}^\ast G_m(m_b)C_3   \nonumber \\ &-(C_4+C_6)V_{tb}V_{tD}^\ast[G_m(m_c)+G_m(m_b)].
	\end{align}
	Taking advantage of the effective operators in Eq.(\ref{effoper}),  we obtain the factorization formulae for  NLO amplitudes  of  $\bar B_q \to PP$ decays as follows
	\begin{eqnarray}
		\mathcal{M}^{LV}_{\rm NLO}&=&{-3 C^2_F M^4}/{\sqrt{6}}\int dx_1 dx_2 dx_3\int b_1db_1 b_2db_2\phi_B(x_1,b_1) \{\mathbf{H}_L({\bf x},{\bf b})\mathbf{C}^{\rm eff}_{D,1}(t_n)\nonumber\\
		&\times&[(x_2-x_3+1)\Phi^{AA}_{23}+2r_2r_3(x_2-x_3+3)\Phi^{PP}_{23}+2r_2r_3(-x_2+x_3+1)\Phi^{TT}_{23}] E'_{a'}(t_n)\nonumber\\
		&+&\mathbf{H}^\prime_L({\bf x},{\bf b})\mathbf{C}^{\rm eff}_{D,1}(t_n)(x_2-x_3+1)[2r_2r_3(\Phi^{TT}_{23}-\Phi^{PP}_{23})-\Phi^{AA}_{23}] E'_{a'}(t'_n)\}
		\\
		\mathcal{M}^{RV}_{\rm NLO}&=&{-3 C^2_F M^5}/{\sqrt{6}}\int dx_1 dx_2 dx_3\int b_1db_1 b_2db_2\phi_B(x_1,b_1) \{h_{nan}(x_1,x_2,x_3,b_1,b_2)C_{D,2}^{\rm eff}(t_n),\nonumber\\
		&\times&[(2x_2x_3-x_2-x_3+1)\Phi^{AA}_{23}+4r_2r_3(x_2x_3-x_2+1)\Phi^{PP}_{23}+4r_2r_3(-x_2x_3+x_3)\Phi^{TT}_{23} \nonumber\\
		&+&2r_2r_3(x_2+x_3-1)(\Phi^{TP}_{23}+\Phi^{PT}_{23} )] E'_{a'}(t_n)\nonumber\\
		&+&h'_{nan}(x_1,x_2,x_3,b_1,b_2)C_{D,2}^{\rm eff}(t_n')[(2x_2x_3-x_2-x_3+1)(\Phi^{AA}_{23}+4r_2r_3\Phi^{PP}_{23})\nonumber\\
		&+&2r_2r_3(x_2+x_3-1)(\Phi^{TP}_{23}+\Phi^{PT}_{23} )] E'_{a'}(t'_n)\}.
	\end{eqnarray}
	Here $\Phi^{ab}_{23}\equiv \phi^a(x_2) \phi^b (x_3)+\phi^a(1-x_2) \phi^b (1-x_3)$,  where the second term stems from the crossing diagram which are not depicted  Fig. \ref{fig:NLO}.  The NLO Feynman diagrams involve three propagators. In   comparison with the leading order scenario,  a new type of hard function emerges.  We write $\mathbf{H}^{(\prime)}_L({\bf x},{\bf b})=(h^{(\prime)}_{na}(x_1,x_2,x_3,b_1,b_2), h^{(\prime)}_{nan}(x_1,x_2,x_3,b_1,b_2))$,   and the effective Wilson coefficients $\mathbf{C}^{\rm eff}_{D,1}(t_n)=({C}^{\rm eff}_{D,1q}(t_n),{C}^{\rm eff}_{D,1m}(t_n))^T$.  The specific expression for $h^{(\prime)}_{nan}(x_1,x_2,x_3,b_1,b_2)$ is shown in the appendix  \ref{pqcd}. 
	The NLO contribution to the $\bar B_s \to \pi\pi$ decay amplitude is then expressed by 
	\begin{align}
		\mathcal{A}_{\rm NLO}(\bar{B}^0_s\rightarrow \pi^+\pi^-)&=\frac{G_F}{\sqrt{2}}\mathcal{ M}^{LV}_{\rm NLO}\left [\mathbf{C}_{s,1}^{\rm eff}\right]+\frac{G_F}{\sqrt{2}}\mathcal{ M}^{RV}_{\rm NLO} \left[C_{s, 2}^{\rm eff}\right].
	\end{align}
	The relation $\sqrt{2}A_{\rm NLO}(\bar{B}^0_s\rightarrow \pi^0\pi^0)=A_{\rm NLO}(\bar{B}^0_s\rightarrow \pi^+\pi^-)$ still holds .  Similarly, the $\bar B_d\to  K^+K^-$ decay amplitude reads
	\begin{align}
		\mathcal{A}_{\rm NLO}(\bar{B}^0_d\rightarrow K^+K^-)&=\frac{G_F}{\sqrt{2}}\mathcal{ M}^{LV}_{\rm NLO}\left [\mathbf{C}_{d,1}^{\rm eff}\right]+\frac{G_F}{\sqrt{2}}\mathcal{ M}^{RV}_{\rm NLO} \left[C_{d, 2}^{\rm eff}\right].
	\end{align}
	For the $B \to VV $ decays, the NLO contribution also depends  significantly  on the polarization states. If the final states are longitudinally  polarized,  the decay amplitude is very similar to the case where the final states are  pseudo-scalar mesons.  For the transversely polarized amplitude,  however, the  matrix element  $\langle V_2(p_2, \varepsilon_{2\perp}^\ast)V_3(p_3, \varepsilon_{3\perp}^\ast)|g^\ast(p_g,\alpha) g^\ast(\tilde p_g,\beta)\rangle$    should be proportional to $g_\perp^{\beta\nu}\,\varepsilon_{2\perp}^{*}\cdot \varepsilon_{3\perp}^{*} -\varepsilon_{2\perp}^{*\beta}\,\varepsilon_{3\perp}^{*\nu} -\varepsilon_{2\perp}^{*\nu}\,\varepsilon_{3\perp}^{*\beta}$.  Contracting this matrix element with the $\bar B \to g^\ast g^\ast$ transition matrix element  results in a zero contribution.  Consequently, only the longitudinal polarization amplitude is affected by the NLO contribution. 
	Similar to the leading order situation, the longitudinally polarized factorization function $\tilde{\mathcal{ M}}^{LV}_{\rm NLO}$ can be obtained from those in the $B_q\to PP$ decays with  	the  replacement  rule in Eq(\ref{replace})  and $r_i\to m_i/m_{B_q}$. The NLO contribution to the longitudinal $\bar B_q \to VV$ decay amplitude is then expressed as 
	\begin{align}
		\mathcal{A}_{L,\rm NLO}(\bar{B}^0_s\rightarrow \rho^+\rho^-)&=\frac{G_F}{\sqrt{2}} \tilde{\mathcal{ M}}^{LV}_{\rm NLO}\left [\mathbf{C}_{s,1}^{\rm eff}\right]+\frac{G_F}{\sqrt{2}}\tilde{\mathcal{ M}}^{RV}_{\rm NLO} \left[C_{s, 2}^{\rm eff}\right],\nonumber \\
		\mathcal{A}_{L,\rm NLO}(\bar{B}^0_s\rightarrow K^{\ast+}K^{\ast-})&=\frac{G_F}{\sqrt{2}}\tilde{\mathcal{ M}}^{LV}_{\rm NLO}\left [\mathbf{C}_{d,1}^{\rm eff}\right]+\frac{G_F}{\sqrt{2}}\tilde{\mathcal{ M}}^{RV}_{\rm NLO} \left[C_{d, 2}^{\rm eff}\right].
	\end{align}
	The decay amplitude of $B \to \phi\phi$ has the same form as 	$\mathcal{A}_{L,\rm NLO}(\bar{B}^0_s\rightarrow K^{\ast+}K^{\ast-})$.  For $B_s \to \rho^0\rho^0, \omega \omega $ decays, the second equation in Eq(\ref{rhorhorelation}) still holds.

	\section{Numerical Analysis}
	
LCDAs are essential non-perturbative entities within the collinear factorization framework. They are characterized by the fact that they contain only a single variable, namely the momentum fraction. Transverse-momentum-dependent (TMD) wave functions, in contrast, are far more complex than a straightforward extension of LCDAs. This complexity arises from the appearance of light-cone singularities. These light - cone divergences can be regularized through the rapidity of non-light-cone Wilson lines. However, this regularization comes at the cost of the emergence of enhanced large logarithms of rapidity, which must be resummed to minimize the scheme dependence.
From a phenomenological perspective, one approach is to utilize a wave function that has been fitted to experimental data. In this study, we employ the model
	\begin{equation}
		\phi_{B_q}(x,b) = N_{B_q} x^2(1-x)^2 \exp \left[ -\frac{M_{B_q}^2\
			x^2}{2 \omega_b^2} -\frac{1}{2} (\omega_b b)^2 \right],\label{waveb}
	\end{equation}
	with $N_{B_q}$ the normalization factor. The  parameter $\omega_b$ is chosen as: $\omega_b=0.40\pm0.5~\mathrm{GeV}$ for $B$ meson and  $\omega_b=0.50\pm0.5~\mathrm{GeV}$   for $B_s$ meson, which has been fixed  using the rich experimental data.  
	For the wave function of the light meson,  the previous studies have indicate that  the transverse momentum dependent part is not important numerically,  therefore, only the longitudinal momentum fraction dependent part is employed.  For the  pseudo-scalar meson,  we  have
	\begin{eqnarray}
		\phi_{P}^A(x)&=&\frac{3f_{P}}{\sqrt{6}}x(1-x)[1+a^P_1(\mu)C^{3/ 2}_1(t)+a^P_2(\mu)C^{3/ 2}_2(t)],  \nonumber\\
		\phi_{P}^P(x)&=&\frac{f_{P}}{2\sqrt{6}}\bigg[1+3\rho^P(1-3a^P_1+6a_2^P)(1+\ln x)-{3\rho^P\over 2}(1-9a^P_1+18a_2^P)C^{1/2}_1(t)\nonumber \\ & -&3\rho^P(a^P_1-5a_2^P)C^{1/2}_2(t)+30\eta_{3P} C^{1/2}_2(t) -{9\rho^P\over 2}a_2^PC^{1/2}_3(t)\bigg], \nonumber\\
		\phi_{P}^T(x)&=&\frac{f_{P}}{2\sqrt{6}}\bigg\{ -t \bigg[1+{\rho^P\over 2}(2-15a^P_1+30a_2^P)+3\rho^P(1-3a^P_1+6a_2^P)\ln x\nonumber \\ &+&\rho^P(3a^P_1-{15\over 2}a_2^P)C^{3/2}_1(t)+{1\over2}(10\eta_{3P}+3\rho a_2^P)C^{3/2}_2(t)\bigg]\nonumber\\
		&+&x(1-x)\bigg[6\rho^P(3a^P_1-{15\over 2}a_2^P)+5(10\eta_{3P}+{3\rho^P}a_2^P)C^{3/2}_1(t)\bigg] \nonumber \\ &+&3(1-x)\rho^P(1-3a^P_1+6a_2^P)\bigg\},
	\end{eqnarray}
	where $t=2x-1$, $\rho^\pi=0$ and $\rho^K=m_s^2/m_K^2$.  For the longitudinally polarized vector meson, the corresponding LCDAs read
	\begin{eqnarray}
		\phi_{V}(x)&=&\frac{3f^\parallel_{V}}{\sqrt{6}}x(1-x)[1+a^{V,\parallel}_1(\mu)C^{3/ 2}_1(t)+a^{V,\parallel}_2(\mu)C^{3/ 2}_2(t)],  \nonumber\\
		\phi_{V}^t(x)&=&\frac{3f^\perp_{V}}{2\sqrt{6}}t\bigg[C^{1/2}_1(t)+a^{V,\perp}_{1}C^{1/2}_2(t)+a^{V,\perp}_{2}C^{1/2}_3(t)\bigg], \nonumber\\
		\phi_{V}^s(x)&=&\frac{3f^\perp_{V}}{2\sqrt{6}}x(1-x)\bigg[1+{a^{V,\perp}_{1}\over 3}C^{1/2}_2(t)+{a^{V,\perp}_{2}\over 6}C^{1/2}_3(t)\bigg].
	\end{eqnarray}
	We do not list the LCDAs of transversely polarized vector meson here, since it does not participate in the NLO amplitudes.  In the numerical calculation of the leading order contribution, we take advantage of the same LCDAs with \cite{Chai:2022ptk}. The decay constants and  Gegenbauer moments of the light mesons, as well as the
	CKM matrix elements  and other parameters,  are listed in Table, \ref{inputs}.   The CKM matrix elements are taken from Particle Data Group. 
	\begin{table}
		\centering
		\renewcommand{\arraystretch}{2.0}
		{		\begin{tabular}{c|c|c|c|c|c}
				\hline
				\hline
				$f_\pi$(GeV) & \,\, $f_K$(GeV)  \,\, &  \,\, $f^\parallel_{\rho^\pm}(f^\parallel_{\rho^0})$(GeV)   \,\, &  \,\, $f^\parallel_\omega$(GeV) \,\,&  \,\, $f^\parallel_{K^\ast}$(GeV) \,\, &  $f^\parallel_\phi$(GeV) \,\, \\
				$0.130$ &$0.156$ &$0.210(0.212)$  & $0.197 $&  \,\, 0.204\,\, & $0.233 $ \\ \hline
				$m_{0\pi}$(GeV) \,\, &  $m_{0K}$(GeV)&  $f^\perp_{\rho^\pm}(f^\perp_{\rho^0})$  (GeV) \,\, &  \,\, $f^\perp_\omega$(GeV) \,\, &  \,\, $f^\perp_{K^\ast}$(GeV) \,\,&  $f^\perp_\phi$(GeV) \,\, \\
				1.400 & 1.892 &0.144(0.146) & 0.162 &0.159 & 0.191  \\ \hline
				$a^\pi_2$ &$a_2^{\rho,\parallel}$& $a_2^{\rho,\perp}$ & $a_2^{\omega,\parallel}$& $a_1^{\omega,\perp}$ & $a_2^{\phi,\parallel}$ \\
				$0.250\pm0.150$ &$0.180\pm0.037$ &$0.137\pm0.030$  & $0.150\pm0.120$& $0.140\pm0.120 $& $0.230\pm0.080 $\\
				\hline
				$a^K_1$ &$a_2^K$& $a_1^{K^\ast,\parallel}$ & $a_1^{K^\ast,\perp}$& $a_2^{K^\ast,\parallel}$ & $a_2^{K^\ast,\perp}$ \\
				$0.076\pm0.004$ &$0.221\pm0.082$ &$0.060\pm0.040$  & $0.040\pm0.030 $& $0.160\pm0.090 $& $0.100\pm0.080 $\\
				\hline
				$a_2^{\phi,\perp}$ &$m_s$(GeV)& $|V_{ud}|$ & $|V_{us}|$& $10^3|V_{td}|$ & $10^3|V_{ts}|$ \\
				$0.140\pm0.070$ &$0.130\pm0.015$ &$0.974$  & $0.224 $& $8.6\pm0.2$& $41.5\pm0.9 $
				\\   			\hline
				$	10^3|V_{ub}| $&$|V_{tb}|$&$\alpha$& $\gamma$&$\eta_{3\pi}$&$\eta_{3K}$  \\
				$3.82\pm0.20$ &$1.010\pm0.027$ &$(84.1^{+4.5}_{-3.8})^\circ$  & $(65.7\pm3.0)^\circ $&$0.025\pm 0.008$ & $0.015\pm 0.005$
				\\ \hline\hline
			\end{tabular}
		}
		\renewcommand{\arraystretch}{1.0}
		\caption{The inputs of parameters in the light meson LCDAs and CKM matrix elements.}
		\label{inputs}
	\end{table}
	
	In order to develop a transparent understanding of the numerical feature for the  identified enhancement mechanism,
	we present in Table \ref{table for the predicted amplitudes} the obtained results of 
	$\mathcal{A}_{\rm LO}(\bar{B}^0_s\rightarrow \pi^+\pi^-, \rho^+\rho^-)$ and  $\mathcal{A}_{\rm NLO}(\bar{B}^0_s\rightarrow \pi^+\pi^-, \rho^+\rho^-)$ for the weak annihilation decays $\bar B_s \to \pi^{+}  \pi^{-}, \rho^+\rho^-$.  It is obvious that the NLO amplitudes contain large imaginary parts,  which makes the NLO contribution comparable to the leading order one,  the ratio of  their moduli  can reach $35\% \sim 40\%$, which is in line as our expectation.   The imaginary part of the NLO amplitudes have opposite sign to  that of leading order amplitude,  thus there exists a cancellation between them. As the result,  the NLO contribution does not cause a significant change in the modulus  of the  total amplitudes much.  Such as  the $\bar B_s \to \pi^{+}  \pi^{-}$ process,  the   modulus of the  total amplitude remains  almost unchanged after including the NLO contribution.   Therefore, the prediction of the branching ratio for this process will not change significantly either, which is consistent with the conclusion in Ref. \cite{Lu:2022kos}, where only the contributions of the leading- twist light-cone distribution amplitudes (LCDA) of light mesons were considered.    In  Table  \ref{table for the predicted branching ratios},  the branching ratios of the pure annihilation type B decay modes including  NLO corrections  are presented. For sake of a comparison, the leading order predictions are given in the parentheses. As expected, in almost all processes, the new results are very close to the previous predictions. In Table \ref{table for the predicted amplitudes}, we also list the experimental results of the processes considered in this paper. It can be seen that our predictions are in good agreement with the experimental data.   For the $B_q \to VV$ processes, we also give the predicted values of the longitudinal fraction. It is found that the pure annihilation-type B-meson decays are all dominated by longitudinal polarization. This is because the transverse polarization amplitudes are severely suppressed. Although the NLO corrections can only contribute to the longitudinal polarization amplitude, they do not cause a significant change in the longitudinal polarization fraction. 
	
	\begin{table}
		\centering
		\renewcommand{\arraystretch}{2.0}
		{		\begin{tabular}{c|c|c|c|c}
				\hline
				\hline
				&	$\mathcal{A}_{\rm LO}$ & \,\, $\mathcal{A}_{\rm NLO}$  \,\, &  \,\, $
				{|\mathcal{A}_{\rm NLO}|\over| \mathcal{A}_{\rm LO}|}$   \,\, &  \,\,${|\mathcal{A}_{\rm LO+NLO}|\over| \mathcal{A}_{\rm LO}|}$\,\, \\
				\hline
				$\bar{B}^0_s\rightarrow \pi^+\pi^-$ &$4.97+3.97\,i$ &$1.26 - 2.07 \, i$ & $0.38$  & $1.02 $ \\
				$\bar{B}^0_s\rightarrow \rho_L^+\rho_L^-$ &$7.06+11.5\,i$& $3.83 - 2.90  \, i$ & $0.36$ & $1.02$  \\
				\hline
				\hline
				&	$\Delta \delta|_{\rm twist-2}$ & \,\, $\Delta \delta|_{\rm twist-3}$  \,\, &  \,\, $\Delta \delta|_{\rm total}$  \,\, &  \,\,$\Delta \delta|_{\rm LO}$ \,\, \\
				\hline
				$\bar{B}^0_s\rightarrow \pi^+\pi^-$& $35.8^\circ$   &  $33.6^\circ$  &  $38.0^\circ$ &$7.2^\circ$  \\	
				$\bar{B}^0_s\rightarrow \rho_L^+\rho_L^-$& $-166.0^\circ$   & $-112.7^\circ$   & $30.3^\circ$  &$-8.9^\circ$  \\
				\hline
				\hline
			\end{tabular}
		}
		\renewcommand{\arraystretch}{1.0}
		\caption{Numerical predictions of the decay amplitudes $\mathcal{A}_{\rm LO}$
			and $\mathcal{A}_{\rm NLO}$  and strong phase difference between tree amplitude and penguin amplitude  for $\bar B_s \to \pi^{+}  \pi^{-}, \rho^+ \rho^-$
			decays}
		\label{table for the predicted amplitudes}
	\end{table}

	As mentioned before, the dominant NLO contribution from the charm-loop diagrams generates the considerable cancellation of  the imaginary part  of leading power contribution, 	in consequence,  the NLO effect considered in this work provides an important source of  the strong phases of the weak-annihilation non-leptonic $\bar B_q$-meson decay amplitudes, which can lead to considerable impact on the CP violating observables.  We first demonstrate the influence 	of the high-order corrections on the strong phase. The decay amplitudes can be decomposed as  $\mathcal{A}\sim V_{ub}V_{uD}^\ast A_u+V_{tb}V_{tD}^\ast A_t(D=d, s)$, then the strong phase of  $A_{u,t}$ is written by $\delta_{u,t}$ respectively, and $\Delta \delta\equiv \delta_{u}-\delta_{t}$.  The direct CP violation is proportional to the sine of the difference in the strong phases and the sine of the difference in the weak phases of the two amplitudes, i.e.,  ${\cal A}_{\rm CP}^{\rm dir}\sim \sin \Delta \delta\sin \Delta\phi$ ,  it  is obviously very sensitive to  $\Delta \delta$.  We present the  difference in the strong phases of $\bar{B}^0_s\rightarrow \pi^+\pi^-, \rho^+\rho^-$ decays in Table \ref{table for the predicted amplitudes}, where $\Delta \delta$ from  twist-2 LCDAs and twist-3 LCDAs of the light meson are listed for a comparison.  It is found that both of them can generate sizable strong phase difference,   which significantly modifies the leading order result. 	As a result,  the predicted $\sin \Delta\delta$ is over 0.5 for both  $\bar{B}^0_s\rightarrow \pi^+\pi^-$  and $ \bar{B}^0_s\rightarrow\rho^+\rho^-$ decays. Therefore, in these processes, objective direct CP violation can be anticipated.
	
	The time-dependent CP asymmetries for the neutral $\bar B_q$-meson
	decaying into CP eigenstates are defined by
	\begin{eqnarray}
		{\cal A}_{\rm CP}(t) &=& \frac{\Gamma (\bar B_q \to M_1 M_2)-\Gamma (B_q \to M_1 M_2)}
		{\Gamma (\bar B_q \to M_1 M_2)+\Gamma (B_q \to M_1 M_2)}
		\nonumber \\
		&=& - \frac{{\cal A}_{\rm CP}^{\rm dir} \, \cos (\Delta m_q \, t) + {\cal A}_{\rm CP}^{\rm mix} \, \sin (\Delta m_q \, t)}
		{\cosh(\Delta \Gamma_q \, t /2) + {\cal A}_{\Delta \Gamma} \,  \sinh(\Delta \Gamma_q \, t /2)} \,,
		\hspace{0.5 cm}
	\end{eqnarray}
	where the observables  ${\cal A}_{\rm CP}^{\rm dir}$, ${\cal A}_{\rm CP}^{\rm mix}$
	and ${\cal A}_{\Delta \Gamma}$ can be found in
	\cite{ParticleDataGroup:2020ssz,Nierste:2009wg,Proceedings:2001rdi},
	satisfying the exact algebra relation
	$|{\cal A}_{\rm CP}^{\rm dir}|^2 + |{\cal A}_{\rm CP}^{\rm mix}|^2 + |{\cal A}_{\Delta \Gamma}|^2 = 1$.
	The  predictions for the two independent CP asymmetries
	(${\cal A}_{\rm CP}^{\rm dir}$ and ${\cal A}_{\rm CP}^{\rm mix}$) are 
	displayed in Table \ref{table for the predicted CP asymmetries},  it is
	evidently revealed that  NLO QCD correction significantly enhances
	both  ${\cal A}_{\rm CP}^{\rm dir}$ and ${\cal A}_{\rm CP}^{\rm mix}$
	of the decay processes studied in this paper.   The direct CP asymmetries of $B_s \to \pi\pi, \rho\rho, \omega\omega$ is predicted to about $-5\%$, which need to be tested in the future experiment.  The absolute value of our prediction is smaller than that in \cite{Lu:2022kos},  which shows the differences in the sources of strong phases between the PQCD and QCDF approaches. Notably,  the predicted  direct CP asymmetry  in $B \to \phi_L \phi_L$ approaches $-40\%$, and strong phase difference is totally from the NLO contribution. 	Compared to direct CP voilation ${\cal A}_{\rm CP}^{\rm dir}$, the observable ${\cal A}_{\rm CP}^{\rm mix}$  receives 
	more significant corrections  from NLO contribution,  since it is proportional to the imaginary part of the decay amplitudes.
	An exception is  the  $\bar B_s \to \rho_{L}  \, \omega_{L} $ decays,  where ${\cal A}_{\rm CP}^{\rm dir}$ and ${\cal A}_{\rm CP}^{\rm mix}$ are kept unchanged. 
	This is because in this process, the final-state isospin is 1, and the production through the two-gluon process violates the conservation of isospin, i.e., the matrix element
	$ \langle \rho_{L}(p)  \, \omega_{L}(q) | g^{\ast}(p_g, \alpha) \,  g^{\ast}(\tilde{p}_g, \beta) \rangle$ vanishes if the tiny $\rho-\omega$ mixing effect is neglected.
	The uncertainties of the branching ratios and CP asymmetries are also present in Table \ref{table for the predicted branching ratios} and Table \ref{table for the predicted CP asymmetries}.  The main sources of the uncertainty is the Gegenbauer moments of the light mesons and the parameter $\omega$ in the B meson wave functions.  The uncertainties from difference sources are added in quadrature.

	\begin{table}
		\centering
		\renewcommand{\arraystretch}{2.0}
		{\begin{tabular}{c|c|c|c}
				\hline
				\hline
				& \,\, $10^6\mathcal{B}|_{\rm Theory}$  \,\, & \,\, $10^6\mathcal{B}|_{\rm Exp. }$  \,\,  &  \,\, $f_L$   \,\,  \\
				\hline
				\,\, $\bar B_s \to \pi^{+} \, \pi^{-},$  \,\, & $0.39^{+0.19}_{-0.18}$ ($0.36 ^{+0.21}_{-0.18}$)  & $0.72\pm 0.10$
				& -   \\
				\,\, $\bar B_s \to  \pi^{0} \, \pi^{0}$  \,\, & $0.19^{+0.10}_{-0.09}$ ($0.18 ^{+0.11}_{-0.10}$)  &  $<7.7$
				& $1.0$ ($1.0$)   \\
				$\bar B_s \to \, \rho^{0}  \, \rho^{0} $ & $0.89 ^{+0.19}_{-0.17}$ ($0.82^{+0.19}_{-0.16}$) & $<320$
				& $\sim1.0$ ($\sim1.0$)   \\
				$\bar B_s \to \rho^{+} \, \rho^{-}, \,$ & $1.71^{+0.36}_{-0.32}$ ($1.58 ^{+0.36}_{-0.30}$) & -
				& $\sim1.0$ ($\sim1.0$)   \\
				$\bar B_s \to \omega \, \omega$ &  $0.62^{+0.27}_{-0.25}$ ($0.55 ^{+0.31}_{-0.25}$)  & -
				& $\sim1.0$ ($\sim1.0$)   \\
				$\bar B_s \to \rho  \, \omega$ & $\sim 0$  ($\sim 0$ ) & -
				& $\sim1.0$ ($\sim1.0$)   \\
				\hline
				$\bar B_d \to K^{+} \, K^{-}$ &  $0.12^{+0.05}_{-0.03}$ ($0.11 ^{+0.04}_{-0.03}$) & $0.078\pm0.015$
				& -  \\
				$\bar B_d \to K^{\ast +}  \, K^{\ast -} $ & $0.14^{+0.06}_{-0.05}$ ($0.12^{+0.06}_{-0.04}$) &  $<0.4$
				& $\sim1.0$ ($\sim1.0$)   \\
				$\bar B_d \to \phi \, \phi $ & $0.029^{+0.010}_{-0.010}$ ($0.015 ^{+0.007}_{-0.005}$) & $<0.027$
				& $0.99$ ($0.97$)   \\
				\hline
				\hline
			\end{tabular}
		}
		\renewcommand{\arraystretch}{1.0}
		\caption{Theory predictions of the branching ratio ${\cal B}$(in unites of $10^{-6}$)
			and the percentage of the longitudinal polarizations $f_L$ 
			for the weak annihilation $\bar B_q \to P P, \, V_L V_L$ decay processes,
			where we have included in parentheses the corresponding LO QCD results for a comparison.}
	\label{table for the predicted branching ratios}
\end{table}

\begin{table}
	\centering
	\renewcommand{\arraystretch}{2.0}
	{\begin{tabular}{c|c|c}
		\hline
		\hline
		& \,\, ${\cal A}_{\rm CP}^{\rm dir}$  \,\, &  \,\, ${\cal A}_{\rm CP}^{\rm mix}$   \,\,  \\
		\hline
		\,\, $\bar B_s \to \pi^{+} \, \pi^{-}, \, \pi^{0} \, \pi^{0}$  \,\, & $-6.0^{+1.1}_{-2.5}$ ($-3.6 ^{+1.8}_{-3.1}$)
		& $-4.2^{+21.4}_{-9.0}$ ($35.9^{+15.6}_{-11.2}$)   \\
		$\bar B_s \to \rho^{+}_{L} \, \rho^{-}_{L} , \, \rho^{0}_{L}  \, \rho^{0}_{L} $ & $-4.2^{+0.7}_{-0.5}$ ($-1.9 ^{+0.7}_{-0.7}$)
		&  $-4.3^{+21.5}_{-9.0}$ ($35.9^{+15.6}_{-11.2}$) \\
		$\bar B_s \to \omega_{L}  \, \omega_{L} $ &  $-4.6^{+1.2}_{-2.0}$ ($-2.6 ^{+1.6}_{-2.4}$)
		&  $-3.8^{+21.8}_{-9.7}$ ($35.9^{+15.6}_{-11.2}$) \\
		$\bar B_s \to \rho_{L}  \, \omega_{L} $ & $0.0 \pm 0.0$ ($0.0 \pm 0.0$ )
		&  $-71.0^{+6.3}_{-5.4}$ ($-71.0^{+6.3}_{-5.4}$) \\
		\hline
		$\bar B_d \to K^{+} \, K^{-}$ &  $41.6^{+12.5}_{-12.3}$ ($38.7 ^{+13.2}_{-12.2}$)
		&  $-2.2^{+19.1}_{-26.4}$ ($-47.0^{+15.7}_{-18.8}$) \\
		$\bar B_d \to K^{\ast +}_{L}  \, K^{\ast -}_{L} $ & $36.7^{+16.0}_{-9.5}$ ($25.4 ^{+17.4}_{-11.1}$)
		&  $-1.4^{+19.7}_{-26.9}$ ($-47.0^{+15.7}_{-18.8}$) \\
		$\bar B_d \to \phi_{L}  \, \phi_{L} $ & $-39.7^{+6.1}_{-8.4}$ ($0.0$)
		& $27.8^{+5.7}_{-25.9}$  ($0.0 \pm 0.0$) \\
		\hline
		\hline
	\end{tabular}
}
	\renewcommand{\arraystretch}{1.0}
	\caption{Theory predictions of the CP asymmetries ${\cal A}_{\rm CP}^{\rm dir}$
		and  ${\cal A}_{\rm CP}^{\rm mix}$ (in unites of $10^{-2}$)
		for the weak annihilation $\bar B_q \to P P, \, V_L V_L$ decay processes,
		where we have included in parentheses the corresponding LO QCD results for a comparison.}
\label{table for the predicted CP asymmetries}
\end{table}

%
\section{Conclusions}
%

Nonleptonic B-meson decays provide an ideal place  for unraveling the fundamental nature of CP violation. In this context, weak annihilation contributions are not merely significant but absolutely crucial,  they play an indispensable role in driving the CP violation processes observed in nonleptonic B-meson decays.
In this paper,  we  are endeavored to perform an improved study on the pure annihilation type charmless hardonic B-decays,  including $B^0\to K^+K^-, K^{\ast+}K^{\ast-},\phi\phi$  and $B_s\to \pi\pi, \rho\rho, \rho\omega,\omega\omega$, within the framework of PQCD approach.   Compared to the previous studies,   we  included a type of enhanced contribution to the various decay processes at next-to-leading order in the strong coupling constant $\alpha_s$.    The enhanced contribution arises from the quark loop contractions of the four quark operators , which might be accompanied  by large Wilson coefficients and/or the multiplication CKM factors.   We only considered the $B\to PP, VV$ decay processes  and do not pay attention to $B \to PV$ modes since the quark-loop diagram will not contribute to these processes.   In PQCD approach,  there is no endpoint singularity in the convolution of  the hard kernel and the wave functions of initial and final state mesons,  and  we have employed both twist-2 and two-particle twist-3  distribution amplitudes in our calculation.  The NLO amplitudes contain large imaginary parts with opposite sign to  that of leading order amplitude,  and the cancellation effect leads to the fact that the NLO contribution cannot have a significant impact on the branching ratios as well as longitudinal polarization fractions(for  $B \to VV$ modes ) of the considered processes. Meanwhile, the NLO effect considered in this work provides an important source of  the strong phases of the weak-annihilation non-leptonic $\bar B_q$-meson decay amplitudes, which can lead to considerable impact on the CP violating observables such as  the CP asymmetry ${\cal A}_{\rm CP}^{\rm dir}$ and mixing induced CP asymmetry ${\cal A}_{\rm CP}^{\rm mix}$. The numerical result evidently revealed that  NLO QCD correction significantly enhances both  ${\cal A}_{\rm CP}^{\rm dir}$ and ${\cal A}_{\rm CP}^{\rm mix}$ of the decay processes studied in this paper,  and the observable ${\cal A}_{\rm CP}^{\rm mix}$  receives  more significant corrections  from NLO contribution,  since it is proportional to the imaginary part of the decay amplitudes.

%

\section*{Acknowledgements}
This work is partly supported  by the  National Natural Science Foundation of China  with
Grant No. 12175218 and 12435004,  and the Natural Science Foundation of Shandong with Grant No.   ZR2024MA076,  ZR2022ZD26  and   ZR2022MA035.


\appendix
\section*{Appendix}
\addcontentsline{toc}{section}{Appendices}

\renewcommand{\theequation}{\Alph{section}.\arabic{equation}}
\renewcommand{\thetable}{\Alph{table}}
\setcounter{equation}{0}
\setcounter{section}{0}
\setcounter{table}{0}

\section{PQCD functions}\label{pqcd}

The evolution factors $E^{(\prime)}_a$ and $E^{(\prime)}_{a^\prime}$
entering in the expressions for the factorization formulae are
given by
\begin{align}
	E'_{a}(t)=\alpha _s(t)\exp[-S_B(t)-S_2(t)-S_3(t)],\nonumber\\
	E'_{a'}(t)=\alpha^2 _s(t)\exp[-S_B(t)-S_2(t)-S_3(t)],
\end{align}
where we have not included the  threshold resummation factor since it cannot bring about significant impact on the contribution from annihilation diagrams. 
The hard scales are chosen as
 \begin{eqnarray}
	t_c&=&\mbox{max}\{\sqrt{\bar x_3}M_{B_q},1/b_2,1/b_3\},\\
	t_c^\prime
	&=&\mbox{max}\{\sqrt {x_2}M_{B_q},1/b_2,1/b_3\},\\
	t_d&=&\mbox{max}\{\sqrt {x_2\bar x_3}M_{B_q},
	\sqrt{1-(\bar x_2 -x_1)x_3}M_{B_q},1/b_1,1/b_2\},\\
	t_d^\prime&=&\mbox{max}\{\sqrt{x_2\bar x_3}M_{B_q},\sqrt{x_3\bar x_2}M_{B_q}, \sqrt{|x_1-x_2|\bar x_3}M_{B_s},1/b_1,1/b_2\},
	\\
	t_n&=&\mbox{max}\{\sqrt {x_2\bar x_3}M_{B_q}, \sqrt {x_2\bar x_3}M_{B_q},
	\sqrt{1-(\bar x_2-x_1)x_3}M_{B_q},1/b_1,1/b_2\},\\
	t_n^\prime&=&\mbox{max}\{\sqrt{x_2\bar x_3}M_{B_q},\sqrt{x_3\bar x_2}M_{B_q}, \sqrt{|x_1-x_2|(1-x_3)}M_{B_q},1/b_1,1/b_2\}.
\end{eqnarray} The Sudakov factor associated in $B_q$ and light meson wave functions read respectively as 
\begin{eqnarray}
&&S_{B}(x_1, b_1, \mu)= s\left(x_1 \frac{m_B}{\sqrt{2}}, b_1\right) + s_q(b_1, \mu) \,, 
\label{eq:sudakov-B}\\
&&S_{i}(x_i, \bar{x}_i, b_i, \mu)= s\left(x_i \frac{m_B}{\sqrt{2}}, b_i\right) + s\left(\bar{x}_i \frac{m_B}{\sqrt{2}}, b_i\right) +  s_q(b_i, \mu) \,,
\label{eq:sudakov-M}
\end{eqnarray}
where $\bar x_i=1-x_i$. The factor $s(Q,b)$  collects the resummation of the double logarithms containing the transverse momentum of partons inside the initial- and final- state mesons,
\begin{eqnarray}
s(\xi Q,b) &=& \frac{A^{(1)}}{2\, \beta_1} \, \hat{q} \, \ln \left( \frac{\hat{q}}{\hat{b}} \right) 
+ \frac{A^{(2)}}{4 \, \beta_1^2} \left( \frac{\hat{q}}{\hat{b}} - 1 \right) - \frac{A^{(1)}}{2 \, \beta_1} \left( \hat{q} - \hat{b} \right) \nonumber \\
&-& \frac{A^{(1)} \, \beta_2 \, \hat{q}}{4\, \beta_1^3} \left[ \frac{\ln(2\, \hat{b}) + 1}{\hat{b}} - \frac{\ln(2 \, \hat{q}) + 1}{\hat{q}} \right] 
- \left[ \frac{A^{(2)}}{4\,\beta_1^2} - \frac{A^{(1)}}{4 \,\beta_1} \, \ln\left(\frac{e^{2\,\gamma_E - 1}}{2} \right) \right] \ln\left(\frac{\hat{q}}{\hat{b}}\right) \nonumber \\
&-&\frac{A^{(1)} \, \beta_2}{8 \, \beta_1^3} \left[ \ln^2(2 \, \hat{b}) - \ln^2(2\, \hat{q}) \right]  
- \frac{A^{(1)} \, \beta_2}{8\,\beta_1^3} \, \ln\left(\frac{e^{2\,\gamma_E - 1}}{2}\right) 
\left[\frac{\ln(2\, \hat{b}) + 1}{\hat{b}} - \frac{\ln(2 \, \hat{q}) + 1}{\hat{q}} \right] \nonumber \\
&-&\frac{A^{(2)} \, \beta_2}{16 \, \beta_1^4} \left[ \frac{2 \ln(2 \, \hat{q}) + 3}{\hat{q}} - \frac{2 \ln(2 \, \hat{b}) + 3}{\hat{b}}\right] 
- \frac{A^{(2)} \, \beta_2}{16 \, \beta_1^4}  \, \frac{\hat{q} - \hat{b}}{\hat{b}^2} \left[ 2\, \ln(2 \, \hat{b}) + 1 \right] \nonumber \\
&-& \frac{A^{(2)} \, \beta_2^2}{432 \, \beta_1^6} \, \frac{\hat{b} - \hat{q}}{\hat{b}^3} \left[ 9 \, \ln^2(2\,\hat{b}) + 6 \, \ln(2\hat{b}) + 2 \right]  \nonumber \\
&-& \frac{A^{(2)} \, \beta_2^2}{1728 \, \beta_1^6} 
\left[ \frac{18 \, \ln^2(2\,\hat{b}) + 30 \, \ln(2\hat{b}) + 19}{\hat{b}^2} - \frac{18 \, \ln^2(2\,\hat{q}) + 30 \, \ln(2\hat{q}) + 19}{\hat{b}^2} \right] \,.
\label{sudakov}
\end{eqnarray}
The abbreviated variables are 
\begin{eqnarray}
\hat{q} \equiv \ln\left(\frac{\xi Q}{\sqrt{2}\Lambda}\right) \,, \,\,\,\,\,\, \hat{b} \equiv \ln\left(\frac{1}{b \Lambda}\right) \,, 
\end{eqnarray}
and the coefficients $A^{(i)}$ and $\beta_i$ are
\begin{eqnarray}
&&A^{(1)} = \frac{4}{3} \,, \,\,\,\,\,\, 
A^{(2)} = \frac{67}{9} - \frac{\pi^2}{3} - \frac{10}{27} \, n_f + \frac{8}{3} \, \beta_1 \, \ln\left(\frac{e^{\gamma_E}}{2}\right) \,,\nonumber \\
&&\beta_1 = \frac{33 - 2 \, n_f}{12} \,, \,\,\,\,\,\, \beta_2 = \frac{152 - 19 \, n_f}{24} \,.
\end{eqnarray}
The factor $s_q(b, \mu)$ arises from the resummation of the single logarithms in the quark self-energy correction
\begin{eqnarray}
s_q(b_1, \mu) = \frac{5}{3} \int_{1/b_1}^\mu \frac{d \bar{\mu}}{\bar{\mu}} \, \gamma_q(g(\bar{\mu})) \,, \quad
s_q(b_i, \mu) = 2 \int_{1/b_i}^\mu \frac{d \bar{\mu}}{\bar{\mu}} \, \gamma_q(g(\bar{\mu})),
\label{eq:sudakov-sq}
\end{eqnarray}
with the quark anomaly dimension $\gamma_q = - \alpha_s(\mu)/\pi$.

The hard functions $h_i$ in decay amplitudes are given by
\begin{align}
&h_{na}(x_1,x_2,x_3,b_1,b_2)=\frac{\pi i}{2}\left[\theta(b_2-b_1)J_0(\sqrt{x_2\bar x_3}Mb_1)H_0 ^{(1)}(\sqrt{x_2\bar x_3}Mb_2)\right.\nonumber\\
&\left.+\theta(b_1-b_2)J_0(\sqrt{x_2\bar x_3}Mb_2)H_0 ^{(1)}(\sqrt{x_2\bar x_3}Mb_1)\right]K_0(\sqrt{1-(\bar x_2-x_1)x_3}Mb_1)
\end{align}
\begin{align}
h'_{na}(x_1,x_2,x_3,b_1,b_2)&=\frac{\pi i}{2}\left[\theta(b_2-b_1)J_0(\sqrt{x_2\bar x_3}Mb_1)H_0 ^{(1)}(\sqrt{x_2\bar x_3}Mb_2)\right.\nonumber\\
&\left.+\theta(b_1-b_2)J_0(\sqrt{x_2\bar x_3}Mb_2)H_0 ^{(1)}(\sqrt{x_2\bar x_3}Mb_1)\right]\nonumber\\
&\times\left\{
\begin{aligned}
	\frac{\pi i}{2} &H_0 ^{(1)}[\sqrt{(x_2-x_1)\bar x_3}Mb_1] ~~~x_1-x_2\textless 0\\
	&K_0[\sqrt{(x_1-x_2)\bar x_3}Mb_1] ~~~~~ x_1-x_2\textgreater 0
\end{aligned}
\right.
\end{align}
\begin{align}
h_{nan}(x_1,x_2,x_3,b_1,b_2)&=\frac{\pi i}{2}\frac{1}{(x_2-x_3)M^2}K_0(\sqrt{1-(\bar x_2-x_1)x_3}Mb_1)\nonumber\\
&\times\left\{ \left[\theta(b_2-b_1)J_0(\sqrt{\bar x_2 x_3}Mb_1)H_0 ^{(1)}(\sqrt{\bar x_2x_3}Mb_2)\right.\right.\nonumber\\
&\left.\left.+\theta(b_1-b_2)J_0(\sqrt{\bar x_2x_3}Mb_2)H_0 ^{(1)}(\sqrt{\bar x_2x_3}Mb_1)\right]-(x_2\leftrightarrow x_3)\right\}
\end{align}
\begin{align}
h'_{nan}(x_1,x_2,x_3,b_1,b_2)&=\frac{\pi i}{2}\frac{1}{(x_2-x_3)M^2}
\left\{ \left[\theta(b_2-b_1)J_0(\sqrt{\bar x_2x_3}Mb_1)H_0 ^{(1)}(\sqrt{\bar x_2x_3}Mb_2)\right.\right.\nonumber\\
&\left.\left.+\theta(b_1-b_2)J_0(\sqrt{\bar x_2x_3}Mb_2)H_0 ^{(1)}(\sqrt{\bar x_2x_3}Mb_1)\right]
-(x_2\leftrightarrow x_3)\right\}\nonumber\\
&\times\left\{
\begin{aligned}
	\frac{\pi i}{2} &H_0 ^{(1)}[\sqrt{(x_2-x_1)\bar x_3}Mb_1] ~~~x_1-x_2\textless 0\\
	&K_0[\sqrt{(x_1-x_2)\bar x_3}Mb_1] ~~~~~ x_1-x_2\textgreater 0
\end{aligned}
\right.
\end{align}

\section{Pure annihilation type $B \to VV$ decay amplitudes}\label{vvamp}

The specific expression for the LO amplitudes of  each pure annihilation decay processes can be written by
\begin{eqnarray}
\mathcal{A}_{i,{\rm LO}}(\bar B_{s}^0\to\rho^{+}\rho^{-}) &=&  \frac{G_F}{\sqrt{2}}
V_{ub}V_{us}^{*} \Big\{ f_{B_s}F_{\rm LO}^{LL,i}\left[a_2\right] +
\mathcal{M}_{\rm LO}^{LL,i}[C_2] \Big\}
\nonumber \\
&&- \frac{G_F}{\sqrt{2}} V_{tb}V_{ts}^{*}\bigg \{ f_{B_s}
F_{\rm LO}^{LL,i}\left[2a_{3}
+\frac{1}{2}a_{9}\right]
+f_{B_s}F_{\rm LO}^{LR,i}\left[2a_{5} +\frac{1}{2}a_{7} \right]
\nonumber \\
&& +\mathcal{M}_{\rm LO}^{LL,i}\left[2C_{4}+\frac{1}{2}C_{10}\right]
+\mathcal{M}_{\rm LO}^{SP,i}\left[2C_{6}+\frac{1}{2}C_{8}\right]\bigg \},
\nonumber \\
\sqrt{2}	\mathcal{A}^i(\bar B_{s}^0\to\rho^{0}\rho^{0}) &=& \sqrt{2} 	\mathcal{A}^i(\bar B_{s}^0\to \omega\omega)=
\mathcal{A}_{i,{\rm LO}}(\bar B_{s}^0\to\rho^{+}\rho^{-}) \label{rhorhorelation}
\end{eqnarray}
\begin{eqnarray}
2	\mathcal{A}^i(\bar B_{s}^0\to\rho^{0}\omega) &=& \frac{G_F}{\sqrt{2}}
V_{ub}V_{us}^{*}
\Big\{ f_{B_s}F_{\rm LO}^{LL,i}\left[a_{2}\right] +  \mathcal{M}_{\rm LO}^{LL,i}[C_{2}] \Big\}
- \frac{G_F}{\sqrt{2}}  V_{tb}V_{ts}^{*}\Big\{ f_{B_s}
F_{ann}^{LL,i}\left[\frac{3}{2}a_9 \right]  \nonumber
\\
&& + f_{B_s}
F_{\rm LO}^{LR,i}\left[\frac{3}{2}a_7 \right]
+\mathcal{M}_{\rm LO}^{LL,i}\left[\frac{3}{2}C_{10}\right]
+\mathcal{M}_{\rm LO}^{SP,i}\left[\frac{3}{2}C_8\right]\Big\} +\left [\rho^0
\leftrightarrow \omega\right ].
\end{eqnarray}
\begin{eqnarray}
\mathcal{A}_{i,{\rm LO}}(\bar B_{d}^0\to K^{\ast +}K^{\ast -}) &=&  \frac{G_F}{\sqrt{2}}
V_{ub}V_{ud}^{*} \Big\{ f_{B_d}F_{\rm LO}^{LL,i}\left[a_2\right] +
\mathcal{M}_{\rm LO}^{LL,i}[C_2] \Big\}
\nonumber \\
&&- \frac{G_F}{\sqrt{2}} V_{tb}V_{td}^{*}\bigg \{ f_{B_d}
F_{\rm LO}^{LL,i}\left[2a_{3}
+\frac{1}{2}a_{9}\right]
+f_{B_d}F_{\rm LO}^{LR,i}\left[2a_{5} +\frac{1}{2}a_{7} \right]
\nonumber \\
&& +\mathcal{M}_{\rm LO}^{LL,i}\left[2C_{4}+\frac{1}{2}C_{10}\right]
+\mathcal{M}_{\rm LO}^{SP,i}\left[2C_{6}+\frac{1}{2}C_{8}\right]\bigg \},
\nonumber \\
\mathcal{A}^i(\bar B_{d}^0\to\phi\phi) &=&- \frac{G_F}{\sqrt{2}} V_{tb}V_{td}^{*}\bigg \{ f_{B_d}
F_{\rm LO}^{LL,i}\left[2a_{3}
+\frac{1}{2}a_{9}\right]
+f_{B_d}F_{\rm LO}^{LR,i}\left[2a_{5} +\frac{1}{2}a_{7} \right]
\nonumber \\
&& +\mathcal{M}_{\rm LO}^{LL,i}\left[2C_{4}+\frac{1}{2}C_{10}\right]
+\mathcal{M}_{\rm LO}^{SP,i}\left[2C_{6}+\frac{1}{2}C_{8}\right]\bigg \},
\end{eqnarray}
where $i=L,N,T$ . 
The longitudinal polarization amplitudes $\mathcal{A}_L$ can be obtained from those in the $B_q\to PP$ decays with
the following replacement in the distribution amplitudes:
\begin{eqnarray}
\phi_{2(3)}^A(x) \rightarrow \phi_{2(3)}(x), \,\,\,
\phi_{2(3)}^P(x) \rightarrow (-)\phi_{2(3)}^s(x),  \,\,\,
\phi_{2(3)}^T(x) \rightarrow (-)\phi_{2(3)}^t(x),
\label{replace}\end{eqnarray}
in addition, the ratios $r_i\to m_i/m_{B_q}$ with $m_{i}$ being vector meson mass.
On the contrary to longitudinal polarization, 
Fig.\ref{fig:LO} (a) and (b) can contribute to normal polarization and transverse polarization amplitudes,  and  the results are collected as follows
\begin{eqnarray}
F_{\rm LO}^{LL(R),N}(a_i)	&=& -8\pi C_Fm_{B_q}^4r_2
r_3\int^1_0dx_2dx_3\int^\infty_0b_2db_2b_3db_3
\Big\{E_a(t_c)a_i(t_c)h_a(x_2,1-x_3,b_2,b_3)
\nonumber\\
&&\left[(2-x_3)\left(\phi_2^v\phi_3^v+\phi_2^a\phi_3^a\right)
+x_3(\phi_2^v\phi_3^a+\phi_2^a\phi_3^v)\right]
-	E_a(t_c')a_i(t_c')h_a(1-x_3,x_2,b_3,b_2)\nonumber\\
&&[(1+x_2)
(\phi_2^v\phi_3^v+\phi_2^a\phi_3^a)	
-(1-x_2)(\phi_2^v\phi_3^a+\phi_2^a\phi_3^v)]
\Big\}.
\nonumber \\
F_{\rm LO}^{LL(R),T}(a_i)
&=& \pm16\pi C_Fm_{B_q}^4r_2
r_3\int^1_0dx_2dx_3\int^\infty_0b_2db_2b_3db_3 \Big\{
E_a(t_c)a_i(t_c)h_a(x_2,1-x_3,b_2,b_3)\nonumber\\
&&[x_3(\phi_2^v\phi_3^v+\phi_2^a\phi_3^a)+(2-x_3)(\phi_2^v\phi_3^a+\phi_2^a\phi_3^v)]
+E_a(t_c')a_i(t_c')h_a(1-x_3,x_2,b_3,b_2)		\nonumber\\
&&[(1-x_2)
(\phi_2^v\phi_3^v+\phi_2^a\phi_3^a)	-(1+x_2)(\phi_2^v\phi_3^a+\phi_2^a\phi_3^v)]
\Big\}.
\end{eqnarray}
For the non-factorizable
annihilation diagrams shown in Fig.\ref{fig:LO} (c) and (d),
we have
\begin{eqnarray}
\mathcal{M}_{\rm LO}^{LL(SP),N}(a_i)	&=&-64\pi C_Fm_{B_q}^4r_2 r_3/\sqrt
{6}\int^1_0dx_1dx_2dx_3\int^\infty_0b_1db_2b_2db_2\phi_{B_q}(x_1,b_1)\nonumber \\ &&[\phi_2^v\phi_3^vx+\phi_2^a\phi_3^a]
E_a'(t_d)a_i(t_d)h_{na}(x_1,x_2,x_3,b_1,b_2),
\nonumber \\
\mathcal{M}_{\rm LO}^{LL(SP),T}(a_i)
&=&\mp 128 \pi C_Fm_{B_q}^4r_2 r_3/\sqrt
{6}\int^1_0dx_1dx_2dx_3\int^\infty_0b_1db_2b_2db_2\phi_{B_q}(x_1,b_1)\nonumber \\ &&[\phi_2^v\phi_3^+\phi_2^a\phi_3^v]
E_a'(t_d)a_i(t_d)h_{na}(x_1,x_2,x_3,b_1,b_2),
\end{eqnarray}

\bibliographystyle{apsrev4-1}
\bibliography{References}

\end{document}